\documentclass[12pt]{revtex4}

\usepackage{graphicx}

\newcommand{\be}{\begin{equation}}
\newcommand{\ee}{\end{equation}}
\newcommand{\bea}{\begin{eqnarray}}
\newcommand{\eea}{\end{eqnarray}}
\newcommand{\br}{\hskip .25cm/\hskip -.25cm}

\newcommand{\nonu}{\nonumber\\}

\begin{document}

\pagestyle{empty}
\begin{flushright}
{CERN-PH-TH/2009-003}\\
\end{flushright}
\vspace*{5mm}
\begin{center}
{\large {\bf Do Three Dimensions tell us Anything\\
 about a Theory of Everything?}} \\
\vspace*{1cm}
{\bf Jean Alexandre$^1$}, {\bf John~Ellis$^{1,2}$} and {\bf Nikolaos E. Mavromatos$^1$} \\
\vspace{0.3cm}
$^1$ Department of Physics, King's College, London, England \\
$^2$ Theory Division, Physics Department, CERN, CH-1211 Geneva 23, Switzerland \\

\vspace*{2cm}
{\bf ABSTRACT} \\
\end{center}
\vspace*{5mm}
\noindent
It has been conjectured that four-dimensional ${\cal N}=8$ supergravity
may provide a suitable framework for a `Theory of Everything', if its
composite SU(8) gauge fields become dynamical. We point out that
supersymmetric three-dimensional coset field theories
motivated by lattice models provide toy
laboratories for aspects of this conjecture. They feature dynamical composite supermultiplets made
of constituent holons and spinons.
We show how these models may be extended to include ${\cal N}=1$ and ${\cal N}=2$
supersymmetry, enabling dynamical conjectures to be verified more rigorously.
We point out some special features of these three-dimensional
models, and mention open questions about their relevance to the dynamics
of ${\cal N}=8$ supergravity.
\vspace*{3cm}
\noindent

\begin{flushleft} CERN-PH-TH/2009-003 \\
January 2009\\
\end{flushleft}
\vfill\eject

\setcounter{page}{1}
\pagestyle{plain}

\pagebreak

\section{Introduction}

Once upon a time, four-dimensional ${\cal N}=8$ supergravity was touted as a
possible `Theory of Everything'~\cite{TOE}. This suggestion was sparked by the
observation that ${\cal N}=8$ supergravity has a hidden, composite SU(8) gauge
structure~\cite{Nicolai}. The interacting scalar fields of ${\cal N}=8$ supergravity have a non-compact
$E_{7(7)}$ structure with 133 components that are reduced by the 63
generators of the SU(8) symmetry to the 70 physical, on-shell scalar fields.
Attention was drawn by the realization that SU(8) is a gauge symmetry large
enough to include a flavour symmetry as well as an SU(5) grand-unification group.
This motivated the suggestion that the composite SU(8) gauge fields might
become physical particles~\cite{TOE}, as a result of some unspecified dynamical
mechanism. Already in the first papers on this subject, attention was drawn to
two-dimensional coset models in which the composite gauge fields became
dynamical as a result of infrared singularities~\cite{Nicolai}.

However, the suggestion that ${\cal N}=8$ supergravity was a promising framework for
a `Theory of Everything' rapidly became unfashionable, for several reasons. One was
that no convincing mechanism for making the gauge fields
dynamical came to mind, and another was that the
SU(8) gauge symmetry was thought to be anomalous~\cite{Ferrara}. However, the big blow was the
realization that ten-dimensional $E_8 \times E_8$ and SO(32) heterotic string theories
had no anomalies, and could provide suitable frameworks for grand unification,
after compactification to four dimensions~\cite{heterotic}. Subsequently, many promising
four-dimensional models were derived from string theory as understanding of
string dynamics deepened. This deepening understanding led to the realization
that all string theories are linked, and to the discovery that they are related to
eleven-dimensional supergravity~\cite{Mtheory}. In turn, this theory
may be related to ${\cal N}=8$ supergravity
by compactification to four dimensions~\cite{d11}. However, there is no simple limit in which
string theory can be compactified to four-dimensional ${\cal N}=8$ supergravity without
additional low-mass fields~\cite{Green}.

Interest in ${\cal N}=8$ supergravity has recently revived, motivated by the realization
that it is very well-behaved in the ultraviolet, and may well be finite in perturbation
theory~\cite{finite}. This development has brought back to the collective consciousness a
`forgotten' paper by Marcus~\cite{Marcus}, in which he showed that the SU(8) gauge
symmetry of ${\cal N}=8$ supergravity is, in fact, anomaly-free. This paper had been
overlooked in the euphoria surrounding string theory but now, when coupled with
the good ultraviolet behaviour of ${\cal N}=8$ supergravity, it motivates a re-examination
of the possibility that this might be a promising road towards a `Theory of
Everything'. This possibility is not in necessarily conflict with the candidacy of string theory as
the `Theory of Everything', at least as long as the relation of string theory to ${\cal N}=8$
supergravity remains to be elucidated.

In this paper we review the relevant aspects of three-dimensional lattice models
that may serve as inspirations for modelling
the dynamics of ${\cal N}=8$ supergravity. In particular,
we display the role of dynamical composite supermultiplets made of constituent holons and spinons.
As we show explicitly, supersymmetry is an inessential complication, in the sense that it does
not alter the nature of the infrared behaviour. On the other hand, the elevation of
${\cal N}=1$ supersymmetry to ${\cal N}=2$ does enable dynamical results to be placed on a more
rigorous basis. We finish by recalling some of the limitations
of the three-dimensional models, and highlighting some of the questions that arise
before the existence of a similar mechanism in ${\cal N}=8$ supergravity could be addressed.

\section{2+1-Dimensional Condensed-Matter Models with Dynamical Gauge Bosons
\label{sec:2}}

\begin{figure}[t]
\centering
\includegraphics[width=7cm]{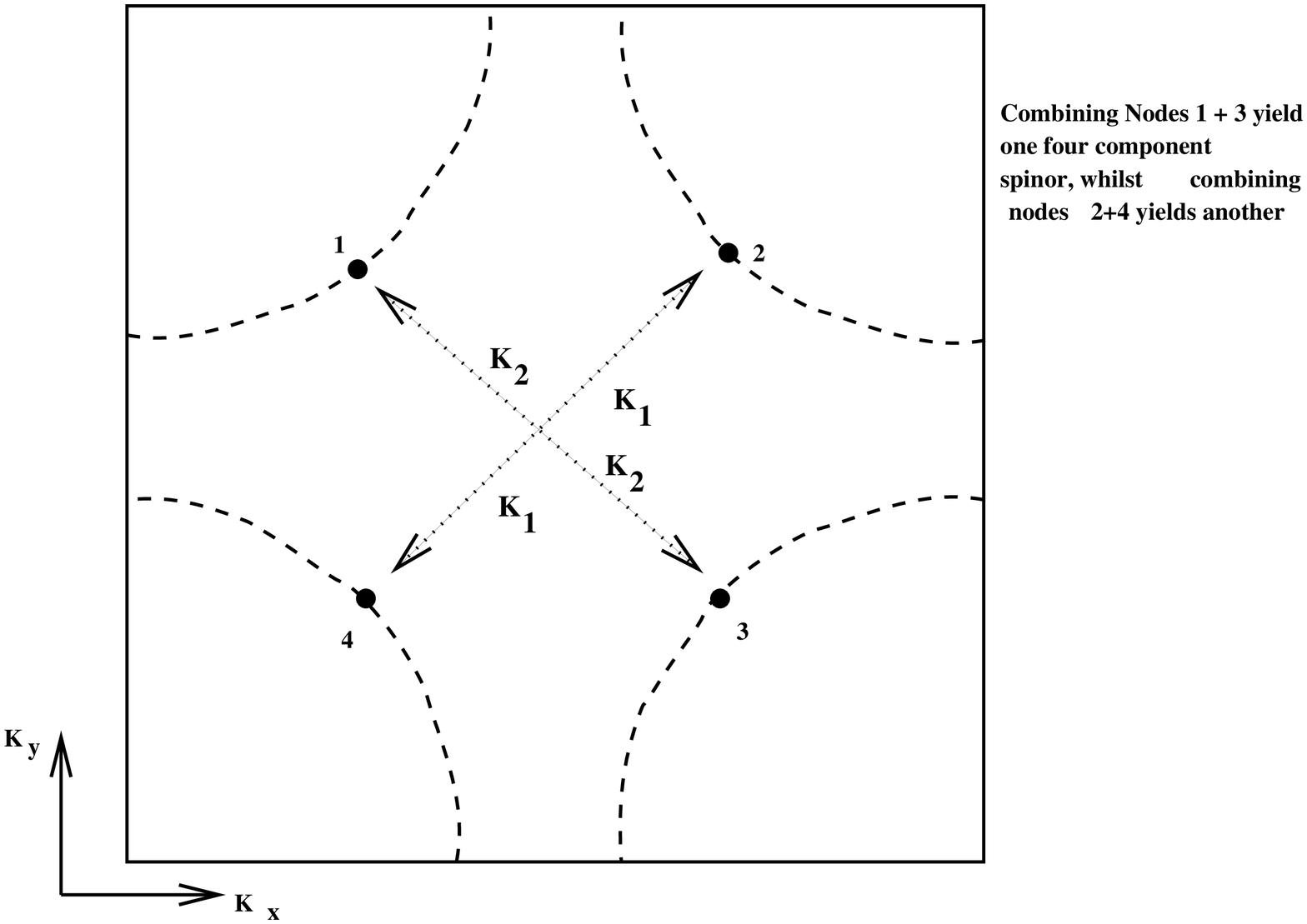}
\hfill \includegraphics[width=8cm]{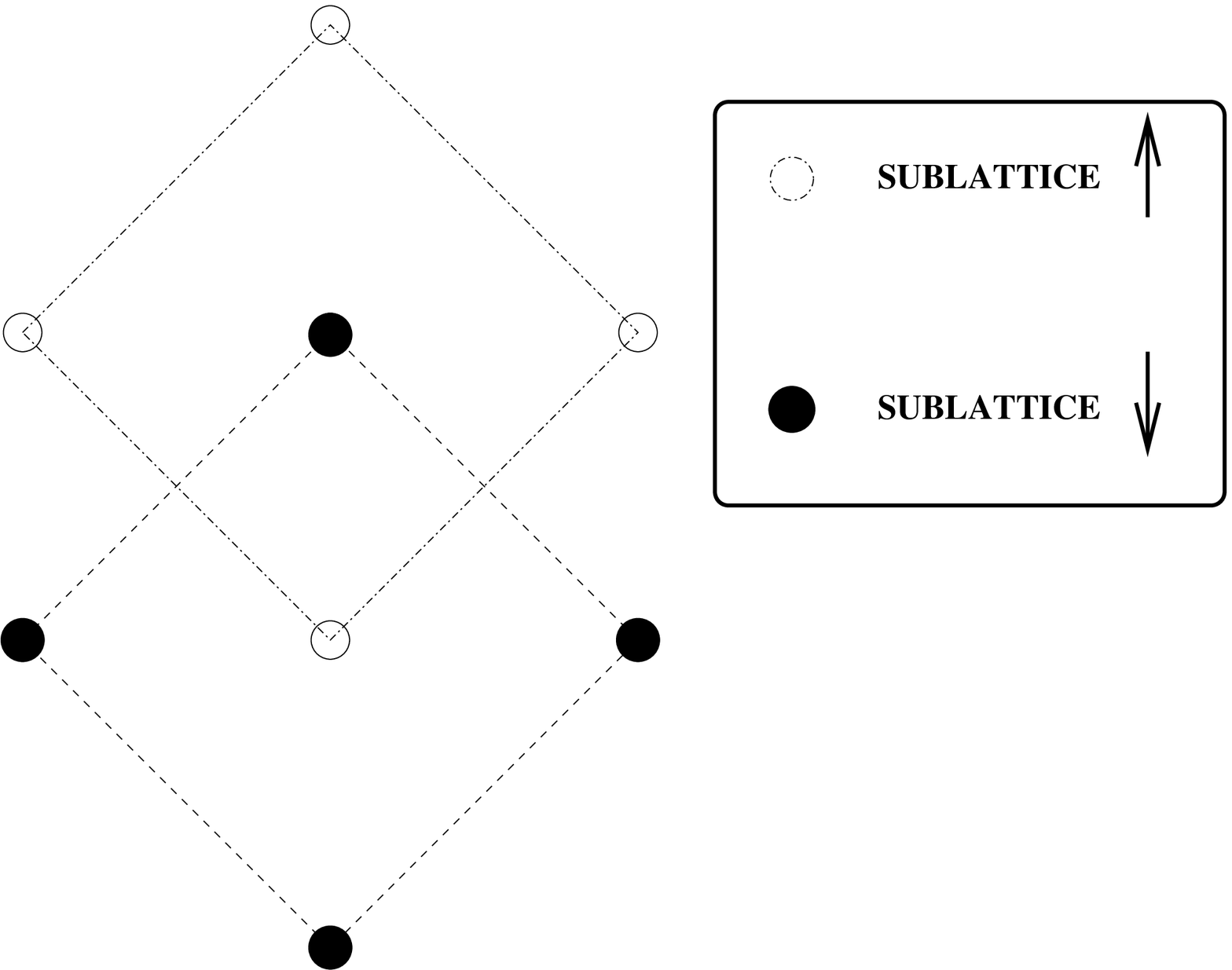}
{\small \caption{\it Left: The Fermi surface of underdoped cuprates consists of four nodes,
as indicated by the dashed lines. The continuum effective theory
may be obtained in a standard way by linearisation about such nodes,
which leads, at the constituent level, to two flavours of
four-component Dirac spinors for the holon degrees of freedom.
These flavours are obtained
by combining the nodes along the diagonal lines, as indicated in the figure:
nodes 1 + 3 yield one flavour, and nodes 2 + 4 yield the other.
Right: A simple antiferromagnetic sublattice structure, which can be used
to define local (spin) SU(2) gauge groups, playing the r\^ole of `colour'. There are two coloured
four-component spinors in this construction, each obtained by taking the continuum limit in a
sublattice. The dynamical breaking of the local SU(2) with mass generation for fermions implies suppression of inter-sublattice hopping in such models.
\label{fig:nodes}}}
\end{figure}

The understanding of field theories in three space-time dimensions has largely been
driven by the impetus to study condensed-matter systems formulated on a lattice.
These field theories become relativistic near nodes of the Fermi surface, see
Fig.~\ref{fig:nodes}, and may exhibit dynamical gauge bosons and
supersymmetry, as we discuss below.

A classic model studied with antiferromagnetism and high-temperature
superconductivity has been the  $t-J$ Hamiltonian~\cite{AM,anderson},
which is expressed in terms of
electron creation operators $c^\dagger _{a, i}$ and their conjugates.
As shown in~\cite{farakos,mav}, when inter-sublattice hopping is included, this Hamiltonian can be expressed in terms of field operators $\chi$, whose structure
we discuss below. For example, the hopping term reads
\begin{equation}
H_{hop}=-\sum_{\langle ij \rangle
}t_{ij} c^{\dagger}_{\alpha,i}c_{\alpha,j}=
-\sum_{\langle ij \rangle }t_{ij}
[\chi^{\dagger}_{i,\alpha\gamma}\chi_{j,\gamma\alpha}
+ \chi^{\dagger}_{i,\alpha\gamma}(\sigma_3)_{\gamma\beta}
\chi_{j,\beta\alpha}] ,
\label{Hop7}
\end{equation}
where $i$ denotes a lattice site, $\sigma _3$ is a $2\times 2$ Pauli matrix, and summation
over the spin indices is implied.
To this may be added Heisenberg interaction terms,
which can be written
in the following convenient form~\cite{AM}
\be
 H_{\rm HI}=-\frac{1}{8}J\sum_{<ij>} tr[\chi_i \chi^\dagger _j \chi
_j\chi^\dagger_i].
\label{heisenb}
\ee
The antiferromagnetic nature of such systems translates into the existence of
more than one sublattice (see Fig.~\ref{fig:nodes}), associated with the spin states of the
hole excitations, with intersublattice hopping non-zero but suppressed relative to the hopping
inside a single sublattice. As we review below,
this results in internal gauge degrees of freedom for the hole excitations, which
becomes a fully-fledged dynamical non-Abelian gauge symmetry in the effective field theory.
There is an even number of sublattices or nodes in the Fermi surface of such
systems, and hence an even number of spinor degrees of freedom, which results in parity
remaining unbroken. (We remind the reader that in 2+1 dimensions a single fermion
mass term breaks parity, but this can be restored if there is an even number of fermionic species,
with mass terms in pairs of opposite signs.)

To see this, as discussed in~\cite{farakos,mav}, one may represent the field $\chi$ via a
spin-charge separation Ansatz in the case of
a planar antiferromagnetic lattice with intersublattice
hopping for a particle-hole symmetric formulation away from half-filling:
\begin{equation}
\chi_i \equiv \left(\begin{array}{cc}
{\psi _1} \qquad {\psi _2} \\
 {-\psi _2^\dagger} \qquad {\psi _1^\dagger}\end{array}
\right)_i\left( \begin{array}{cc}
 {z_1} \qquad {-\bar z_2} \\
 {z_2} \qquad {\bar z_1}\end{array} \right)_i \; .
\label{ansatz}
\end{equation}
The fields $z_a$ obey canonical {\it bosonic}
commutation relations, and are associated with the
{\it spin} degrees of freedom (`spinons'), whilst the fields $\psi_a$
are Grassmann variables on the lattice, which obey {\it fermionic}
statistics, and are associated with the electric charge degrees of freedom
(`holons').

The ansatz (\ref{ansatz}) admits a \emph{hidden} non-Abelian local SU(2)
spin symmetry at the constituent level, as discussed in~\cite{farakos,mav}.
This may readily be seen by considering
the invariance of the $\chi$ variable under simultaneous local SU(2)
rotations of the spinon and holon components:
\begin{eqnarray}
&~&\Psi _{i}\longrightarrow \Psi _{i}\;h_{i}\qquad {\rm where}\qquad \Psi
\equiv \left(
\begin{array}{cc}
\psi _{1}\qquad  & \psi _{2} \\
\psi _{2}^{\dagger }\qquad  & -\psi _{1}^{\dagger }%
\end{array}%
\right)   \nonumber \\
&~&Z_{i}\longrightarrow h_{i}^{\dagger }\;Z_{i}\qquad {\rm where}\qquad
Z\equiv \left(
\begin{array}{cc}
z_{1}\qquad  & -\overline{z}_{2} \\
z_{2}\qquad  & \overline{z}_{1}%
\end{array}%
\right)
\end{eqnarray}%
where $h_{i}\in $ SU(2).
In three space-time dimensions, due to the fractional statistics of the
planar excitations, one has an additional Abelian phase rotation, forming a U$_S$(1)
`statistical' Abelian gauge group (which should not be confused with the
U(1) of electromagnetism).
Hence, the full local gauge group of the spin-separation ansatz is SU(2)~$\otimes$~U$_S$(1).
In terms of the spin and charge
excitations appearing in (\ref{ansatz}),
the hopping term in the Hamiltonian (\ref{Hop7}) may be written as
\bea
H_{hop}=-\sum_{\langle ij \rangle }t_{ij}
[{\overline z}_{i,b}{\psi^\dagger}_{i,b,\alpha}
\psi_{j,c,\alpha}z_{j,c}
+ {\overline z}_{i,b}{\psi^\dagger}_{i,b,\alpha}
(\sigma_3)_{\alpha\beta}\psi_{j,c,\beta}
z_{j,c}]
\label{Hop8}
\eea
which has a trivial local SU(2) symmetry.

The Heisenberg interaction term (\ref{heisenb})
can be \emph{linearized} in terms of the fermion bilinears
if one introduces in the path integral
a Hubbard-Stratonovich field $\Delta_{ij}$,
in a standard fashion.
The result of the linearization for the combined hopping and interaction
Hamiltonian is~\cite{farakos,mav} had a Hartree-Fock form:
\begin{eqnarray}
&~&H_{{\rm HF}}=\sum_{\langle ij\rangle }{\rm Tr}\left[ \frac{8}{J}\Delta
_{ij}^{\dagger }\,\Delta _{ji}+\left( -t_{ij}(1+\sigma _{3})+\Delta
_{ij}\right) \Psi _{j}^{\dagger }\langle Z_{j}\,\overline{Z}_{i}\rangle \Psi
_{i}\right] +  \nonumber \\
&~&\sum_{\langle ij\rangle }{\rm Tr}\left[ \overline{Z}_{i}\langle \Psi
_{i}^{\dagger }\left( -t_{ij}(1+\sigma _{3})+\Delta _{ij}\right) \Psi
_{j}\rangle Z_{j}+{\it h.c.}\right] .  \label{hf}
\end{eqnarray}%
This Hamiltonian determines the nature of any
spontaneous symmetry breaking that occurs, and the associated phase structure of the
low-energy theory near the nodes of the Fermi surface.
Using the gauge symmetries of (\ref{ansatz}), we can write
\begin{equation}
\langle Z_{j}\overline{Z}_{i}\rangle \equiv |A_{1}|\,{V}%
_{ij}\,U_{ij}~,~\langle \Psi _{i}^{\dagger }\left( -t_{ij}(1+\sigma
_{3})+\Delta _{ij}\right) \Psi _{j}\rangle \equiv |A_{2}|\,{V}_{ij}\,U_{ij}~,
\label{bilinears}
\end{equation}%
where ${V}\in $ SU(2) and $U\in $ U$_{S}$(1) are group elements. The group
U$_{S}$(1) expresses the fractional statistics of the three-dimensional
excitations of spinons $Z$ and holons $\Psi$.
The fact that apparently gauge non-invariant correlators are non-zero on the
lattice is standard in gauge theories, and does not violate Elitzur's
theorem~\cite{elitzur}, due to the fact that in order to evaluate the
physical correlators one must follow a gauge-fixing procedure, which should
be done prior to any computation. The amplitudes $|A_{1}|$ and $|A_2| \equiv K > 0$ are
considered frozen, which is a standard assumption in the gauge theory
approach to strongly correlated electron systems~\cite{anderson,AM}. The
group elements ${V}$ and $U$
are phases of the above field bilinears and, to a first (mean-field) approximation,
can be considered as {\it composites} of the constituent
spinons $z$ and holons $\psi $. Fluctuations about such ground states can
then be considered by integrating over the constituent fields.

The above lattice action does not contain kinetic terms for
the SU(2) and U$_S$(1) groups, which would take the forms:
\begin{equation}
\sum_{p} \left[\beta_{SU(2)}(1-Tr V_p) + \beta_{U_S(1)}(1-Tr U_p)\right] ,
\label{plaquette}
\end{equation}
where the sum is over the plaquettes $p$ of the lattice, and the coefficients
\begin{equation}
\beta_{U_S(1)} \equiv \beta_1 ~, ~\beta_{SU(2)} \equiv \beta_2 =
4\beta_1~,
\label{couplings}
\end{equation}
are the inverse square couplings of the U$_S$(1) and SU(2) groups,
respectively.
The specific relation between the coefficients of the SU(2) and U$_S$(1)
terms is a consequence
of the appropriate normalizations of the generators
of the groups.
The absence of gauge kinetic terms in our case implies that we are in
the limit of infinitely strong coupling for both gauge groups:
\begin{equation}
\label{strongcoupl}
\beta_{U_S(1)} = \beta_{SU(2)} = 0 \; \leftrightarrow \; g_{U_S(1)}, g_{SU(2)} \to \infty .
\end{equation}
However, as discussed in detail in~\cite{farakos}, integration of the
$Z$-spinon fields results in kinetic terms for gauge fields, which couple to the
spinons $\Psi$  only.
The analysis of~\cite{farakos} assumes that the $Z$-spinons have a mass gap that is
larger than the dynamical mass gap generated by the holon fields through their interactions with the Abelian gauge group, so that one can define the appropriate effective theory of the light
degrees of freedom by integrating out the heavy ones. (We recall that the kinetic terms are the lowest
terms in a derivative expansion, in which terms with more than two derivatives are irrelevant operators in
the infrared, and hence can be ignored when one considers the low-energy continuum
limit of interest here.) In such a case, one obtains~\cite{farakos}:
\begin{equation}
\beta_1 = \frac{1}{J \eta a} ,
\end{equation}
where $a$ is the (sub)lattice spacing, $\eta$ denotes the doping concentration in the sample,
and $J$ is the Heisenberg interaction in the condensed-matter model (\ref{hf}), and group
theory maintains the relationship (\ref{couplings}) between the SU(2) and U$_S$(1)
couplings. The strong-U$_S$(1) coupling corresponds formally to $J \eta a \to \infty$.

We now turn to the dynamical breaking of the SU(2) group~\cite{farakos}, using
a Schwinger-Dyson treatment of dynamical symmetry breaking based on a large-$N$
approximation, in which the SU(2) spin group is replaced by SU(N) with $N$ large.
In this case, the non-Abelian coupling is related to the Abelian one through
\be
\beta_{SU(N)}=2N\beta_{U_S(1)} = 2N \beta_1 .
\label{betarel}
\ee
This implies that, even in the case of strong U$_S$(1) coupling,
 $\beta_1 \rightarrow 0$, the large-$N$ (large-spin)
limit may be implemented in such a way so that $\beta_{SU(N)}$ is finite, and
may be assumed weak.

As discussed in detail in~\cite{farakos} and references therein,
the lattice hamiltonian (\ref{hf}) after integrating out the massive spinon
$Z$ fields, can be expressed in the infrared limit as a
lattice spinor hamiltonian for the holon fields $\Psi$
coupled to the U$_S$(1) and SU(2) gauge fields:
\bea
&~& S=\frac{1}{2}K \sum_{i,\mu}[{\overline \Psi}_i (-\gamma_\mu)
U_{i,\mu}V_{i,\mu} \Psi_{i+\mu}  +  {\overline \Psi}_{i+\mu}
(\gamma _\mu)U^\dagger_{i,\mu}V^\dagger_{i,\mu}
\Psi _i ]  \nonumber  \\
&~& + \beta _1
\sum _{p} (1 - trU_p) + \beta _{SU(N)} \sum _{p} (1-trV_p) ,
\label{effeaction}
\eea
where $\mu =0,1,2$, $U_{i,\mu}=exp(i\theta _{i,\mu})$
represents the
statistical U$_S$(1) gauge field,
$V_{i,\mu}={\rm exp}(i\sigma ^a B_a)$ is the SU(2) gauge field,
and the plaquette terms appear as a result of
the $z$ (spinon) integration, as mentioned previously.
Working in the large $N$-fermion flavour limit ($N$ even sublattices, $N \to \infty$)
and keeping the SU(N) coupling $\beta_{{\rm SU(N)}}$ finite, according to (\ref{betarel}),
the SU(2) gauge field becomes dynamical. The U$_S(1)$ field, on the other hand, is assumed
strongly coupled: $\beta_1 \to \infty$, and hence its kinetic term is absent. The fact
that kinetic terms for the SU(2) gauge fields can be induced in the effective action
by quantum corrections reflects the existence of a non-trivial infrared fixed point in
the (2+1)-dimensional Abelian gauge theory, after the dynamical breaking of the SU(2).
This was discussed in detail in~\cite{papavas}, and has also been discussed in the
${\cal N}=1$ supersymmetric case~\cite{camp}.

After integrating the effective spinon-gauge-field Lagrangian over
the strongly-coupled statistical U$_{S}$(1) dynamical gauge group, one
finds the following effective partition function:
\be
  \int\left[dVd{\overline \Psi}d \Psi\right] exp(-S_{eff}) ,
\label{action3}
\ee
where
\bea
&~&S_{eff} = \beta _2 \sum _{p} (1 -trV_p)
+ \sum _{i,\mu}{\rm ln}I_0(\sqrt{y_{i\mu}}) \nonumber \\
&~&y_{i\mu} \equiv
K^2 {\overline \Psi}_i (- \gamma _\mu)
V_{i\mu}\Psi _{i+ \mu} {\overline \Psi}_{i + \mu}
(\gamma _\mu)V_{i\mu}^\dagger \Psi _i .
\label{action4}
\eea
Here, $K$ is the amplitude $|A_2|$ of the fermionic bilinears in the Hartree-Fock
lattice action (\ref{hf})
and $I_0$ is the zeroth-order Bessel function.
The quantity $y_{i\mu}$ may be written in terms
of the bilinears
\be
M^{(i)}_{ab,\alpha\beta}
\equiv \Psi _{i,b,\beta}{\overline \Psi}_{i,a,\alpha},~~a,b={\rm
colour},~\alpha,\beta={\rm Dirac},~i={\rm lattice~site} ,
\label{mesons}
\ee
with the result:
\be
y_{i\mu} = -K^2tr[M^{(i)}(-\gamma _\mu)V_{i\mu}
M^{(i+\mu)}(\gamma_\mu)V_{i\mu}^\dagger] .
\label{y}
\ee
In the language of particle physics, quantities analogous to the $M^{(i)}$
would represent physical {\it meson} states.
Converting from fermionic to bosonic variables, the low-energy (long-wavelength)
effective action may written as a path integral
in terms of gauge-boson and meson fields~\cite{farakos}
\be
Z=\int [dV dM]exp(-S_{eff}+ \sum_{i}tr{\rm ln}M^{(i)}) ,
\label{effmeson}
\ee
where the meson-dependent term in (\ref{effmeson})
comes from the Jacobian that arises when converting the integral from
fermion to meson variables.

A method was presented in~\cite{farakos} for
identifying the symmetry-breaking patterns of the
gauge theory (\ref{effmeson}), by studying
the
dynamically-generated mass spectrum. The method consists
of expanding $\sum _{i,\mu}{\rm ln}I_0(\sqrt{y_{i,\mu}})$
in powers of $y_{i\mu}$ and concentrating on the lowest
orders, which yield the gauge boson
masses, whilst higher orders describe interactions.
Keeping only the linear term in the expansion
yields
\be
   {\rm ln}I_0(\sqrt{y_{i\mu}}) \simeq
\frac{1}{4}y_{i\mu} =
-\frac{1}{4}K^2 tr[M^{(i)}(-\gamma _\mu)V_{i\mu}
M^{(i+\mu)}(\gamma _\mu)V_{i\mu}^\dagger] .
\label{fkresult}
\ee
It is evident that the pattern
of SU(2) breaking is determined by non-zero VEVs
for the meson matrices $M^{(i)}$.
One has
the following expansion for the
meson states in terms of SU(2)
bilinears~\cite{farakos}:
\bea
 &~& M^{(i)} = {\cal A}_3(i)\sigma_3
+ {\cal A}_1(i)\sigma_1 + {\cal A}_2 (i)\sigma_2 +
{\cal A}_4 (i){\bf 1} + \nonumber \\
&~& i[B_{4,\mu}\gamma ^\mu + B_{1,\mu}(i)\gamma^\mu\sigma_1
+ B_{2,\mu}\gamma^\mu\sigma_2 + B_{3,\mu}\gamma ^\mu \sigma_3 ] ,
\label{mesontripl}
\eea
with $\mu=0,1,2$,
and
\bea
&~&{\cal A}_1 \equiv -i[{\overline \Psi}_1 \Psi_2
- {\overline \Psi}_2 \Psi_1] ,
\qquad {\cal A}_2 \equiv {\overline \Psi}_1 \Psi_2
+ {\overline \Psi}_2 \Psi_1 ,
\qquad {\cal A}_3 \equiv {\overline \Psi}_1\Psi _1
- {\overline \Psi}_2 \Psi_2 ,
\nonumber \\
&~&B_{1\mu} \equiv {\overline \Psi}_1\sigma _\mu \Psi _2 +
{\overline \Psi}_2\sigma_\mu \Psi_1 \quad
B_{2\mu} \equiv i[{\overline \Psi}_1\sigma _\mu \Psi _2 -
{\overline \Psi}_2\sigma_\mu \Psi_1], \quad
B_{3\mu} \equiv {\overline \Psi}_1\sigma _\mu \Psi_1
-{\overline \Psi}_2\sigma _\mu \Psi_2, \nonumber \\
&~&{\cal A}_4 \equiv {\overline \Psi}_1\Psi_1
+ {\overline \Psi}_2\Psi_2, \qquad
B_{4,\mu} \equiv {\overline \Psi}_1\sigma _\mu \Psi_1
+ {\overline \Psi}_2\sigma _\mu \Psi_2~, \qquad \mu=0,1,2 .
\label{2compbilin}
\eea
Here, $A_1, A_2, A_3, B_1, B_2, B_3$ transform as triplets under SU(2),
and $A_4, B_4$ transform as SU(2) singlets,
the  $\gamma_\mu$ are (antihermitean) Dirac (space-time) $2\times 2$ matrices,
and the $\sigma_i$, $i=1,2,3$ are the (hermitean) $2\times 2$
SU(2) Pauli matrices.

Naively, the composite meson fields $A_i$ and $B_i$
transform under a global SU(2) group~\cite{farakos,farak}.
However, because of the Hartree-Fock form of the lattice Hamiltonian (\ref{hf}),
these composites realize a strongly-coupled local (gauged) SU(2) $\otimes $ U$_S$(1) group,
which (following the above discussion) enables the representation of
the SU(2) gauge link variables in terms of the $B_i$ composites~\cite{farakos}:
\be
V_{i\mu} = {\rm exp}(i \sigma^a B_a ) = \cos(|{\bf B}_{i\mu}|) + i{\bf \sigma}.{\bf B}_{i\mu}
\sin(|{\bf B}_{i\mu}|)/|{\bf B}_{i\mu}| ,
\label{linkexpres}
\ee
where $a=1,2,3$ are SU(2) indices, and bold face letters indicate vectors in SU(2)
space. (The reader should bear in mind the trick (\ref{betarel}), of working with large
SU(N) instead of the SU(2) group, so as to guarantee a finite $\beta_{SU(N)}$ coupling, while
working with a strong U$_S$(1) group.)

The VEV of the matrix $<M^{(i)}>=u\sigma_3$
is proportional to the chiral condensate $u$, which is responsible for generating dynamically a
(parity-conserving) mass gap for the holons~\cite{farakos}.
Substituting (\ref{mesontripl})
into (\ref{fkresult}), and performing a naive perturbative expansion
over the fields ${\bf B}$ one finds:
\be
{\rm ln}I_0(\sqrt{y_{i\mu}}) \propto
K^2 u^2
[(B^1_{i\mu})^2 + (B^2_{i\mu})^2]~+~{\rm interaction~terms} .
\label{massiveboson}
\ee
From this it follows that two of the SU(2) gauge bosons,
namely the $B^1$,$B^2$, become massive, with masses
proportional to $K u$, whereas the gauge boson $B^3$ remains massless.
These mass terms break SU(2) down to a U(1) subgroup.

We draw the reader's attention to the
similarity of the above mechanism for symmetry breaking
with the situation
in the adjoint gauge-Higgs model~\cite{adjoint}.
There, the SU(2) symmetry is also broken down to a U(1) subgroup
whenever the constant multiplying the Higgs-gauge interaction
is larger than a critical value.
In our case
the r\^ole of this constant is played by $K^2$,
as can be seen by the formal analogy between
the adjoint-Higgs-gauge interaction terms and (\ref{fkresult}).
In our
approach symmetry
breaking was achieved via
the infinitely-strong U$_S$(1) coupling.
In view of the above analogy with the
adjoint-Higgs model, however,
one may speculate that an interesting
phase diagram for the symmetry breaking of SU(2)
could also emerge due to the $K^2$ coupling,
whatever the U$_S$(1) coupling.

At this point we would like to clarify certain points concerning the above-mentioned models: the lattice systems we examined above
are simplifications of the actual situation encountered in realistic condensed matter systems,
where in some regions of their parameter space the Fermi surface is 
characterized by pockets of finite size rather than points. In such models, the spin-charge separation ansatz is more subtle than the one presented here. The interested reader can find more details on such issue in the recent literature~\cite{recent}, where detailed phase diagrams are presented. However, for our purposes here, we are simply interested in the compositeness aspects of the above 
decomposition of spin and charge degrees of freedom, which are merely used here as a motivation for our particle-physics oriented analysis rather than a rigorous study of realistic condensed matter systems and their properties. In this respect, in what follows we shall make use of the above formalism in order to draw some useful lessons and thus be able to discuss the notion of composite operators and their relevance to N=8 supergravity phenomenology.

We close this Section by mentioning the possibility of extending the above results
to other gauge groups, provided the ansatz (\ref{ansatz}) for spin-charge separation
in the simple SU(2) case that represents the spin degree of freedom in a antiferromagnetic system
is modified accordingly.

\section{Supersymmetric 2+1-Dimensional Condensed-Matter Models with Dynamical Gauge Bosons \label{sec:3}}

\subsection{${\cal N}=1$ supersymmetry}

A further step was taken in~\cite{diamand,ms}, where conditions
were derived under which the continuum low-energy limit of the above lattice Hamiltonian
becomes supersymmetric. First, it was observed
on the basis of an appropriate power counting of the fundamental
spinon and holon degrees of freedom, and taking into account the gauge freedom,
that there are regions in the parameter space of the underlying $t-J$ model which exhibit
dynamical ${\cal N}=1$ supersymmetry between the fundamental constituents of the model,
namely the spinons and holons $z_a, \psi_a$, $a=1,2$ in the Ansatz (\ref{ansatz}).

The authors of~\cite{ms} considered
the lattice $t-J$ models discussed in the previous section, but including
non-nearest-neighbour hopping and interaction terms in the antiferromagnetic
sublattices. Such models have been used in theoretical modeling of
$d$-wave high-temperature superconductors~\cite{[1]}.
The analysis led to the following ${\cal N}=1$ supersymmetric continuum Lagrangian
in terms of component spinon $\psi$ and holon $z$ fields:
\be
L = g^2 \sum_{a=1}^2\left[D_\mu \bar{z}_a D^\mu z_a +
i {\overline \psi}_a \not{D} \psi_\alpha + \bar{f}_a f_a
 + 2i({\overline \eta} \psi_a \bar{z}_a -
{\overline \psi}_a \eta z_a)\right] ,
\label{B.16a}
\ee
where $D_\mu$ denotes the gauge-covariant derivative with respect to the
U$_S$(1) gauge field, and $f_a $ is an auxiliary field.
The field $\eta$ is a Majorana fermion, which is viewed as the supersymmetric partner of the
Abelian U$_S$(1) gauge field, needed to reproduce (as an appropriate Lagrange multiplier
field~\cite{ms}), the CP$^1$ $\sigma$-model constraint, which in superfield notation reads
\be
\sum _{a=1}^2 {\overline \phi}_a \phi_a = 1 ,
\label{B.18}
\ee
with the superfield $\phi_a$ being given by
\be
\phi_a=z_a + \theta\psi_a + \frac{1}{2}\theta^2 f_a .
\label{B.10}
\ee
This contains, for each colour, a complex scalar $z_a$,
a Dirac spinor $\psi_a$, and a complex auxiliary field $f_a$.
We refer the reader to~\cite{ms} for a detailed discussion how this formalism
can actually describe realistic models of doped antiferromagnets in certain regions of parameter
space that guarantee ${\cal N}=1$ supersymmetry, and under what circumstances one can have ${\cal N}=2$ supersymmetric extensions.

For our purposes we note that,
in component form, the constraint (\ref{B.18}) reads for the physical fields:
\be
\sum _{a=1}^{2} |z_a|^2 = 1~~~~\mbox{and}~~~~
\sum _{a=1}^{2}\left(\overline z_a\psi_\alpha+z_a\overline\psi_a\right)=0,
\label{B.8}
\ee
and these results can be generalised to non-Abelian cases, such as the broken
SU(2) case discussed previously.

In addition to supersymmetry between the constituent spinon and holons,
it was suggested in~\cite{ams}, motivated by the Hartree-Fock form (\ref{hf}) and the associated
spinon and holon bilinears (\ref{bilinears}) forming gauged group-link variables,
that composite fields made out of the fundamental constituents $z$ and $\Psi$
exhibit an on-shell ${\cal N}=2$ supersymmetry, in the above-mentioned regions of the parameter
space where the constituent ${\cal N}=1$ supersymmetry exists. Since these composite fields could
constitute observable
excitations of these materials,
this kind of composite supersymmetry could provide a way to obtain some exact results in the phase diagrams
of doped antiferromagnets, and hence high-temperature superconductors.

This induced ${\cal N}=2$ supersymmetric structure in the low-energy composite theory
provides exact results on the phase structure of
doped antiferromagnets in this regime of the corresponding parameter space
because the effective field theory is~\cite{ams} ${\cal N}=2$ scalar QED,
in which extra ${\cal N}=1$ matter multiplets may be present, with non-trivial superpotentials.
The moduli spaces of such theories have been studied extensively in~\cite{intril},
following the pioneering work  of Seiberg and Witten~\cite{seiberg} in four dimensions,
with the result that
there is a non-trivial non-perturbative infrared fixed point,
induced by the matter multiplets. We recall that a (2+1)-dimensional
gauge theory is superrenormalizable, without ultraviolet  divergences,
by elementary power counting. However, it has a non-trivial low-energy (infrared) structure.
There are increasing hints that four-dimensional ${\cal N} = 8$ supergravity theory might also
lack the expected ultraviolet divergences. However, the
low-energy (infrared) structure of the four-dimensional ${\cal N} = 8$
theory remains to be elucidated.

The non-trivial infrared structure of the (2+1)-dimensional theory was interpreted in~\cite{ams}
as indicating a deviation from the trivial Landau-liquid fixed point,
implying several observable properties in the normal phase of the superconductor.
In addition, these theories also exhibited a superconducting phase~\cite{ams,dorey},
associated with a phase in which the SU(2) non-Abelian symmetry
is broken down to an Abelian subgroup.
The superconducting phase corresponds to the Coulomb phase of the Abelian subgroup
(not to be confused with electromagnetic gauge symmetry), in which the corresponding
gauge boson is massless. This leads to an anomalous current-current one loop diagram with a massless
pole which, according to the anomaly mechanism
of~\cite{dorey}, results in superconductivity (according to the Landau criterion)
upon coupling to a real external electromagnetic field.
The Higgs phase of the Abelian subgroup of the original SU(2) gauge group,
in which the gauge field is massive, corresponds to the pseudogap phase of the underdoped
cuprates (in condensed-matter parlance), which occurs for low doping.
In addition, non-perturbative (monopole) configurations have also been discussed~\cite{ams},
in various supersymmetric effective gauge theories, following~\cite{intril},
in an attempt to discuss the formation of domain-wall structures, such as the stripe phase of
high-temperature superconductors, in which the spin and charge are separated in spatial
domains in the underdoped (non-superconducting) regions of the material.
Thus, the emergence of dynamical composite gauge fields in supersymmetric 2+1 lattice models
and non-trivial aspects of its spontaneous breaking by composite scalar fields
can in principle be subjected to experimental tests.

This completes our review of relevant known exact results in condensed-matter systems
in which a composite gauge symmetry is realized dynamically.

\subsection{${\cal N}=2$ supersymmetry}

Having in mind the ${\cal N}=8$ supergravity theory of interest,
we now describe an ${\cal N}=2$ supersymmetric SU(2) coset field theory in 2+1 dimensions,
in which the superpartners of the theory are composite fields
generalizing the previous construction out of fundamental spinon and holon
degrees of freedom. The latter are assumed to belong to ${\cal N}=1$
supermultiplets, but the supersymmetry may be extended to ${\cal N}=2$,
if the gauge condition $\partial^\mu A_\mu=0$ is satisfied.
The authors of~\cite{hlousek} needed this gauge condition in
the Abelian Higgs model in order to identify the gauge field with a topologically-conserved current,
which was at the origin of their elevation of the supersymmetry.
The same condition was found for the $CP^1$ model in~\cite{edelstein},
and was generalized to Yang-Mills theory in~\cite{alexandre}. Its
recurrence suggests that this gauge condition is necessary, model-independently,
for supersymmetry elevation in 2+1 dimensions.

Our starting point is to assume the existence of two complex ${\cal N}=1$ scalar supermultiplets
($i$ is now a flavour indice)
\be\label{phi}
\phi_i=z_i+\theta\psi_i+\frac{1}{2}\theta^2f_i~~~~~~~~i=1,2 \;,
\ee
where the two-component Grassman coordinate $\theta$ is real, and the complex field components
$z_i,\psi_i,f_i$ are respectively the scalar, the two-component fermion and the auxiliary field.
Note that, in 2+1 dimensions, a real two-component fermion represents one degree of freedom (d.o.f.),
such that, in the supermultiplet (\ref{phi}), the complex scalar d.o.f. compensates the complex
fermionic d.o.f., and the role of the
auxiliary field is to balance the numbers of bosonic and fermionic components in the superfield.
We also remind the reader that in 2+1 dimensions there is no chirality condition,
as this would restrict the space-time dependence of the field components.

Based on these fundamental degrees of freedom,
the following ${\cal N}=1$ composite supermultiplets were considered in~\cite{ams}, where
$D_\alpha \equiv \partial/\partial\theta^\alpha+i(\br\partial\theta)_\alpha$ is the
supercovariant derivative:
\begin{itemize}

\item scalar supermultiplets, forming an SU(2) triplet,
\bea\label{Phi}
\Phi^1&=&D^\alpha\phi_1D_\alpha\phi_1\nonu
\Phi^2&=&D^\alpha\phi_2D_\alpha\phi_2\nonu
\Phi^3&=&D^\alpha\phi_1D_\alpha\phi_2;
\eea

\item parity-conserving vector supermultiplets, forming an SU(2) triplet,
\bea\label{V}
V_\alpha^1&=&Re\{\phi_1D_\alpha\phi_2+\phi_2D_\alpha\phi_1\}\nonu
V_\alpha^2&=&Im\{\phi_1D_\alpha\phi_2+\phi_2D_\alpha\phi_1\}\nonu
V_\alpha^3&=&Re\{\phi_1D_\alpha\phi_1-\phi_2D_\alpha\phi_2\};
\eea

\item a parity-violating composite vector supermultiplet, forming an SU(2) singlet,
\be
V_\alpha^4=Re\{\phi_1D_\alpha\phi_1+\phi_2D_\alpha\phi_2\}.
\ee
\end{itemize}

The next step is  is to gather
the ${\cal N}=2$ supermultiplet degrees of freedom into the following complex superfields
\be\label{G}
G^a=Re\{\Phi^a\}+iD^\alpha V^a_\alpha~~~~~~~~a=1,2,3.
\ee
Each of these contains a real scalar field, a gauge field,
a complex two-dimensional gaugino and a real auxiliary field.
Let us denote by $g$ the gauge coupling and by $f^{abc}$ the SU(2) structure constants.
Using the Wess-Zumino gauge for $V^a_\alpha$, it was shown in~\cite{alexandre}
that the following Lagrangian describes the resulting ${\cal N}=2$ supersymmetric SU(2) gauge theory:
\be\label{calL}
{\cal L}=\int d^2\theta\left|D^\alpha G^a
+gf^{abc}\left( G^bV^{c\alpha}+\frac{i}{2}D^\alpha(V^{b\beta}V^c_\beta)\right) \right|^2,
\ee
provided the gauge condition
\be\label{gaugecondition}
\partial^\mu A^a_\mu=0
\ee
holds for any gauge index $a$. This gauge condition was found in~\cite{alexandre}
to be necessary to cancel unwanted
contributions proportional to $\partial^\mu A^a_\mu$, when ${\cal L}$ is expanded in
terms of the field components. Although we exhibit here an SU(2) model,
it is clear that this construction may be extended
to any SU(n) symmetry group, and is not restricted to the $=2$ case presented here.

An important question is how the above-described composite (`meson-like') fields
become dynamical. In Section \ref{sec:2}, we have seen that the SU(2) holon composites,
which are obtained by integrating out a non-dynamical (strongly coupled) U(1), can become
dynamical, as a result of their gauging, which is a direct consequence of the Hartree-Fock lattice hamiltonian (\ref{hf})
and the associated spinon and holon bilinears (\ref{bilinears}).
A similar situation characterises the supersymmetric composites, where again there are
lattice Hamiltonians of Hartree-Fock type, as we discussed extensively in Section \ref{sec:3}.
However there are a few subtleties, which we now outline. Unlike the non-supersymmetric case,
where the mass gaps of the spinon $z$ and holons $\Psi$ were different, in the supersymmetric case one is not allowed to integrate out the spinons, which now are degenerate in mass with the holons. Hence both the groups U$_S$(1) and SU(2) appear strongly coupled
at the constituent level. However, we can apply here again the trick of working with large-N SU(N), instead of the SU(2) group. This implies that, as in the non-supersymmetric case, one can
consider finite couplings for the non-Abelian part, in which case one may demonstrate
that the above-described ${\cal N}=2$ SU(N) supersymmetric composites become dynamical.

Finally, the addition of matter in the fundamental representation necessitates two new ${\cal N}=1$ scalar
supermultiplets, made of composite field components, and containing different degrees of freedom
from those considered so far:
\be\label{Q}
Q_1=\phi_1^2 ,~~~~~~~~~~~~~Q_2=\phi_2^2,
\ee
that we represent as a two-component superfield $Q=(Q_1,Q_2)$. We
note that the matter superfield components must be complex, since they transform
according to the fundamental representation of SU(2).

It is easy to see that the gauge-matter interaction then reads
\be\label{matter-gauge}
\int d^2\theta\left\lbrace \frac{1}{2}\left((D^\alpha Q)^\dagger-igQ^\dagger V^{a\alpha} T^a \right)
\Big(D_\alpha Q+igV^a_\alpha T^a Q\Big)+gQ^\dagger \Phi^a T^a Q\right\rbrace ,
\ee
where the $T^a$ are the SU(2) generators in the fundamental representation.

In this respect, we note that, from the point of view of condensed-matter models, these matter
fields describe interactions external to the two sublattice structures depicted in Fig.~\ref{fig:nodes},
e.g., by coupling the superconducting planes, etc. For our purposes though,
such matter fields may be integrated out, which renders the gauge multiplets dynamical, by inducing
kinetic terms for the gauge fields, in analogy with the $z$-spinon
integration in the non-supersymmetric
models of Section~\ref{sec:2}. In this way one still has formally a strongly-coupled
SU(N) gauge theory, but the quantum corrections renormalise the effective coupling.

To summarize, starting from the two complex ${\cal N}=1$ scalar supermultiplets (\ref{phi}), we have the following construction:
\begin{itemize}
\item Build three complex ${\cal N}=1$ scalar supermultiplets (\ref{Phi}) and three ${\cal N}=1$
vector supermultiplets (\ref{V}), all composed of more elementary degrees of freedom;
\item Gather these degrees of freedom into three ${\cal N}=2$ supermultiplets (\ref{G}), so
as to obtain an ${\cal N}=2$ supersymmetric SU(2) gauge theory, which is
possible if the gauge condition (\ref{gaugecondition}) is imposed;
\item Add the composite matter ${\cal N}=1$ supermultiplets (\ref{Q}) so as to generate an
${\cal N}=2$ supersymmetric SU(2) theory interacting with matter in the fundamental representation.
\end{itemize}
The theory obtained in the Lagrangians (\ref{calL}) and (\ref{matter-gauge}) can of course be
expressed in a more compact way
using the ${\cal N}=2$ superspace formalism, but the point of the present derivation is to start
from ${\cal N}=1$ supersymmetry, using the ${\cal N}=1$ formalism, to arrive at a theory invariant under ${\cal N}=2$ supersymmetry.

As mentioned in the previous Section, one may use exact results on ${\cal N}=2$ gauge theories to
understand the dynamical behaviour of this model.

\section{Summary and Prospects}

We have shown that supersymmetric models in 2+1 dimensions, formulated either at the
lattice level or as coset field theories, exhibit the emergence of dynamical gauge bosons
realizing an SU(N) gauge group.
There are also composite scalar fields whose expectation values may break this
composite gauge symmetry spontaneously to a subgroup that is realized in the Coulomb
phase. Specifically, in the lattice models discussed in Sections \ref{sec:2} and \ref{sec:3},
the underlying constituents are spinons
and holons, and the composite scalars are in the adjoint representation of the gauge
group. In the coset field theories discussed in Section \ref{sec:3}, there may also appear scalar
fields in the fundamental representation of the SU(N) gauge group. By formulating ${\cal N}=2$
supersymmetric versions of these lattice and coset field theories, the derivation of these
results may be placed on a rigorous basis.

An important feature of our models was that the gauge fields were strongly-coupled,
in the sense of not having kinetic terms (i.e., plaquette terms in lattice models).
However, we have argued that, by working in the large $N$-limit, where $N$ is
the number of species (i.e., sublattices, or nodes) which must be even for reasons of
parity conservation, one could formally re-instate a finite value of the SU(N) gauge coupling,
thus making the gauge fields dynamical.

Moreover, by using arguments based on a Hartree-Fock
approximation in the microscopic models, we have
supported the idea that composite gauge fields made out of spinons and holons can,
also in the supersymmetric regions of the parameter space, become dynamical,
leading to full-fledged ${\cal N}=2$ supersymmetric SU(N) gauge theories. In (2+1)-dimensions
these are characterised by a non-trivial infrared fixed-point structure, which constitutes an
additional, more rigorous for inferring the dynamical nature of the gauge fields
from the effects of the quantum corrections. Even if the bare action is characterised by
infinite gauge couplings, quantum loop corrections can generate such terms, thereby
generating the dynamics of the gauge fields. This feature has been seen explicitly in
non-supersymmetric models, by the integration of the spinon (magnon) fields $z$. In
supersymmetric theories, formally one works with  additional matter multiplets, which can be
integrated out, thereby inducing finite (renormalised) gauge kinetic terms and hence rendering
the strongly-coupled gauge theory dynamical.

Although in this work we have dealt explicitly with unitary SU(N) gauge theories, in
which N must be even because of arguments based on the parity invariance that
characterises the antiferromagnetic case,
nevertheless other gauge groups are expected to lead to qualitatively similar conclusions.

These features are exactly what one might like to see~\cite{TOE} in ${\cal N}=8$ supergravity in
3+1 dimensions~\cite{Nicolai}, which has a non-anomalous SU(8)
gauge symmetry~\cite{Marcus}. If this gauge
symmetry would become dynamical, an adjoint scalar multiplet could break it
spontaneously into a rank-7 subgroup such as SU(5) $\times$ U(1)$^3$
or SU(3) $\times$ SU(2) $\times$ U(1)$^3$ , and scalars in fundamental representations
could in principle reduce the rank to the SU(2) $\times$ U(1) gauge group of the
Standard Model. However, infrared behaviour in 3+1 dimensions is different from
that in 2+1 dimensions, and we have no real understanding how the composite
gauge fields might become dynamical in this case. In this connection, we recall that ${\cal N}=8$
supergravity cannot be obtained simply as some limit of string theory without the
appearance of additional low-mass fields, and it is possible that these might play an
essential role.

On the other hand, the results displayed here may be directly applicable to the
2+1-dimensional exceptional supergravity theories that can be obtained by
truncations of ${\cal N}=8$ supergravity in 3+1 dimensions, which have the following
non-compact coset structures: $F_{4(4)}/$USp(6) $\times$ SU(2),
$E_{6(2)}/$SU(6) $\times$ SU(2), $E_{7(-5)}/$SO(12) $\times$ SU(2) and
$E_{8(-24)}/E_7 \times$ SU(2). Since gravity in 2+1 dimensions is topological,
it might be just an `inessential complication' in the understanding of the dynamics
of these theories. They could be more complete laboratories for probing the
dynamics of the 3+1-dimensional $E7(7)/$ SU(8) theory, though they still evade
the tough issue of the infrared behaviour in 3+1 dimensions.

Our work is complementary to the mechanism for making dynamical
the gauge fields of three-dimensional ${\cal N} = 8$ supergravity based on
the maximal supersymmetric effective low-energy theory of multiple M2 branes that was
presented in~\cite{bagger}. There, the relevant non-Abelian gauge fields have no Maxwell term,
but only mixed (parity-conserving) Chern-Simons structures. Our approach is to
suggest that a kinetic term for the gauge field may be generated through renormalization-group
effects in the infrared region of the effective theory, in direct analogy to what happens in planar condensed-matter models on the lattice and their continuum effective theories.
It remains to be seen whether such an approach, which works so well in three space-time
dimensions, can describe correctly the properties of the four-dimensional
${\cal N} = 8$ supergravity theory.

\section*{Acknowledgements}

J.E. thanks Renata Kallosh for rekindling his interest in this subject.
The work of J.E. and N.E.M. was supported in part by the European Union through the Marie Curie Research and Training Network UniverseNet (MRTN-CT-2006-035863), and the work of J.A. was partly supported by The Royal Society (JP080815).

\end{document}